\documentclass{aa}
\usepackage{graphicx}
\newcommand{\lfir}{$L_{\rm FIR}$}
\newcommand{\snrnir}{$SN^{NIR}_r$}
\begin{document}
\title{
The infrared supernova rate in starburst galaxies
		  \thanks{
		  Based on observations collected at the European Southern Observatory,
		  Chile (proposal 66.B-0417),
		  at the Italian Telescopio Nazionale Galileo (TNG) operated on the 
		  island of La Palma by the Centro Galileo Galilei of the INAF 
		  (Istituto Nazionale di Astrofisica),
		  and at the Steward Observatory 61'' telescope}\\
}

%\subtitle{I. Overviewing the $\kappa$-mechanism}

\author{F. Mannucci \inst{1}
          \and R. Maiolino\inst{2,3}
          \and G. Cresci\inst{4}
          \and M. Della Valle\inst{2}
          \and L. Vanzi\inst{5}
          \and F. Ghinassi\inst{6}
          \and V. D. Ivanov\inst{5}
          \and N. M. Nagar\inst{2}
          \and A. Alonso-Herrero\inst{7}
}

\offprints{F. Mannucci}

\institute{Istituto di Radioastronomia, sezione di Firenze,
           Largo E. Fermi 5, I-50125, Firenze, Italy
    \email{filippo@arcetri.astro.it}
    \and INAF, Osservatorio Astrofisico di Arcetri, 
           Largo E. Fermi 5, I-50125, Firenze, Italy
    \and Dipartimento di Astronomia, Univesit\`a di Roma,
           via della vasca navale 86, Roma, Italy
    \and Dipartimento di astronomia, Universit\`a di Firenze, 
	     Largo E. Fermi 5, I-50125, Firenze, Italy
    \and ESO, Ave. Alonso de Cordova 3107, Casilla 19, Santiago 19001, Chile
    \and Telescopio Nazionale Galileo, Aptdo de Correos 565, 
        38700 Santa Cruz de la Palma, Canary Island, Spain.
    \and Steward Observatory, Univ. of Arizona, Tucson, AZ85721, USA.
}

\date{Received May 15, 2002; accepted May 16, 2002}

\abstract{
We report the results of our ongoing search for extincted
supernovae (SNe) at near-infrared wavelengths. 
We have monitored at 2.2$\mu$m
a sample of 46 Luminous Infrared Galaxies and detected 4 SNe.
The number of detections is still small but sufficient
to provide the first estimate of supernova rate at near-infrared wavelengths.
We measure a SN rate of
$SN^{NIR}_r=7.6\pm 3.8$~SNu which is an order of magnitude  larger than 
observed in quiescent galaxies. 
On the other hand, the observed near-infrared rate is
still a factor 3$-$10 smaller than that estimated from the far-infrared
luminosity  of the galaxies. 
Among various possibilities, the most likely scenario is that dust extinction
is so high ($A_V>30$) to obscure most SNe even in the near-IR.

The role of type Ia SNe is also discussed within this context.
We derive the type Ia SN
rate as a function of the stellar mass of the galaxy 
and find a sharp increase
toward galaxies with higher activity of star formation.
This suggests that a significant fraction of type Ia SNe are
associated with young stellar populations. 

Finally, as a by-product, we give the average K-band light curve
of core-collapse SNe based on all the existing data, and
review the relation between 
SN rate and far-infrared luminosity. 

\keywords{ Supernovae:general -- 
           Supernovae:individual:SN1999gd -- 
           Supernovae:individual:SN2000bg --
		   Infrared:stars}
}

\maketitle

%---------------------------------------------------------------------------
\section{Introduction}

The first report by Maiolino et al. (2002) 
on a sample of SNe detected in the near-infrared (NIR)
has opened a new window for the search of these events.
The presence of dust extinction can seriously reduce the measured SN rate in
most of the galaxies, especially in starburst systems where 
dust obscuration is usually larger than in normal galaxies.
This explains the puzzling result by
Richmond et al. (1998) who found starburst and quiescent galaxies to have
the same SN rate. 
A similar result was obtained by Navasardyan et al.
(2001) who looked for SNe in interacting galaxies, finding no evidence for
correlations between SN rate and galaxy properties.

The effects of dust extinction can be vastly reduced by observing in the NIR:
at 2$\mu$m the extinction is about a factor of ten lower than in the optical
V band, allowing for a much deeper penetration into the dusty star-forming
regions. This strategy was originally proposed by van Buren \& Norman (1989).
The first monitoring campaign was conducted a decade ago
and produced the detection of one event, SN~1994bu (van Buren et al., 1994). 
More recently, Grossan et al. (1999) monitored a large number of galaxies
over two years but failed to detect extinguished SNe, 
probably because of their poor
spatial resolution (about 3\arcsec). Bregman, Temi and Rank (2000)
used ISO/ISOCAM to look for mid-infrared spectral signatures of recent SNe 
in the nuclei of 10 nearby galaxies. 
However they failed to detect any emission from SNe, such as the
[NiII] emission line at 6.63$\mu$m which is very bright in SN~1987A
(e.g., Arnett et al., 1989). This is probably due to the fact
that the number of events expected from the far-IR flux
(see section~\ref{sec:snr}) is lower than 0.5.
Lorionov et al. (2002) reported the detection,
while monitoring another event, of SN~2002cv, a bright
type Ia SN showing a high extinction (Meikle \& Mattila, 2002).
A possible SN was also discovered by Mattila et al. (2002) in an archive UKIRT
K-band image.

Most SNe in starburst galaxies are expected to be 
core-collapse events, i.e., to have massive progenitors. The role of type Ia
SNe in these galaxies is the subject of an active debate. 
The progenitors of type Ia SNe are commonly identified with old white dwarfs in
binary systems, but their rate could be enhanced in galaxies with active
star formation
(see, for example, Della Valle and Livio, 1994, and Della Valle \& Panagia,
2003),
as discussed in Sect.~\ref{sec:Ia}. 

Near-infrared monitoring is needed not only to compare 
the SN production in quiescent and starburst 
galaxies, but also to obtain a
complete estimate of the local SN rate to be compared with the rates
at high redshift (Madau et al., 1998; Dahl\'en \& Fransson, 1999,
Sullivan et al. 2000). 
Finally, near-infrared searches
are needed to detect a significant number of SNe in starburst galaxies;
these events are  expected to have peculiar properties: 
the circumstellar medium around them is likely to be
very dense because they occur in molecular clouds 
and the mass-loss rates of the progenitor stars
are expected to be higher because of the increased
metallicity (Vink et al 2001). The effects of this peculiar environment
can be studied only when a significant number of events are discovered
(Bressan et al., 2002).

In 1999 we started a program aimed at monitoring a sample of starburst
galaxies, as discussed in Section 2. 
The most important differences with respect to previous monitoring 
programs are the use of
4-m class telescopes with higher spatial
resolution, providing a deeper limiting
magnitude especially for point sources, and the selection 
of galaxies with higher star-formation rate
in order to have larger expected rates.
This monitoring yielded the detection
of 4 SNe, 2 of which were discovered by our group, 
SN~2001db (Maiolino et al., 2001)
and SN~1999gw (Cresci et al., 2002). SN2001db is the first 
SN detected in the NIR which has received a spectroscopic confirmation. 

%---------------------------------------------------------------------------
\section{The galaxy sample}
\label{sec:sample}

The SN rate (SNr) is roughly proportional to the star-formation rate (SFR).
The best way to have a galaxy sample selected in terms of SFR and  not
biased by
extinction is to use the far-infrared (FIR) luminosity \lfir\ 
as measured with the IRAS 60 and 100$\mu$m bands.
For all but the most dust-free galaxies 
this luminosity is closely proportional to the
SFR as measured from the extinction-corrected optical lines and UV continuum and
from the radio continuum
(see, for example, Mannucci \& Beckwith, 1995; Schearer, 1999; Cram et al.,
1998).

Here we use a definition of FIR luminosity corresponding to the flux:

\begin{equation}
f_{FIR}=1.26\times10^{-14} C \left(2.58\cdot f_{60}+f_{100}\right)
~~~~({\rm W/m^2})~~~ 
\label{eq:fir}
\end{equation}

\noindent
(Helou, 1988),
where $f_{60}$ and $f_{100}$ are the IRAS fluxes at 60 and 100 $\mu$m in Jy
and $C$ is a numerical constant which depends on the $f_{60}/f_{100}$ ratio,
and assumes values between 1.4 and 1.8.
For galaxies with dust temperatures between 20 and 80 K and emissivity between
0 and 2, Eq.~(\ref{eq:fir}) is expected to estimate the total flux between
40 and 120 $\mu$m within 1\%.

We selected galaxies from the various catalogs of IRAS galaxies
(Sanders et al., 1995; Soifer et al., 1987; Kim \& Sanders, 1998)
having \lfir$>10^{11.1}~L_\odot$ which corresponds to
SN rates larger than about 0.4 SN/year (see Section~\ref{sec:snr}).
We chose a limiting distance modulus of 36.5, corresponding to about 200 Mpc,
in order to have a good detection (signal-to-noise ratio of about 10) 
of an average SN (as SN~1980K having M$_{K}=-$18.8 at peak) with no
extinction,
given a typical limiting magnitude of K$\sim$18.5. 
Three galaxies with distance modulus up to 37.3 but with very high 
FIR luminosities were also added to the sample. 
In conclusion, half of the sample is at distances below 100 Mpc.
At 200 Mpc the spatial scale is
200 pc/arcsec, allowing for an easy detection of a point source even within
the central kpc of the galaxies. 

About a third of the galaxies having a spectroscopic classification
suggesting the presence of a Seyfert nucleus, but in all cases the
far-infrared luminosity seems to be dominated by star formation
(see, for example, Corbett et al., 2002, and Thean et al., 2001).
We excluded from the sample
a few galaxies with low luminosity and large distance, and
Mkn~231 as it contains an AGN whose luminosity outshines 
the rest of the galaxy. 

Table~\ref{tab:gallist} lists the 46 observed
galaxies together with coordinates, absolute B band magnitude, 
FIR luminosity, redshift and distance modulus.

\begin{table*}
\caption[]{The galaxy sample. The name, coordinates,
absolute B-band magnitude, log(\lfir/$L_\odot$), and distance modulus
for H$_0$=70 km/sec/Mpc.}
\label{tab:gallist}

\begin{center}
\begin{tabular}{l c c c c c c}
\hline 
Galaxy & \multicolumn{2}{c}{R.A (J2000) DEC.} & $M_{B}$ & \lfir & z &(m$-$M)\\ 
\hline

NGC 34          &00 11 06.5 &$-$12 06 26 &$-$19.42 &11.34 &0.020 & 34.46  \\ 
NGC 232         &00 42 45.6 &$-$23 33 39 &$-$20.44 &11.23 &0.022 & 34.70  \\
MCG +12-02-001  &00 54 03.6 &$+$73 05 12 &$-$21.05 &11.29 &0.016 & 34.03  \\
IC 1623         &01 07 47.2 &$-$17 30 25 &$-$20.56 &11.38 &0.019 & 34.50  \\
UGC 2369        &02 54 01.4 &$+$14 58 14 &$-$21.17 &11.44 &0.031 & 35.49  \\
IRAS 03359+1523 &03 38 46.9 &$+$15 32 55 &$-$20.30 &11.38 &0.035 & 35.76  \\
MCG -03-12-002  &04 21 20.0 &$-$18 48 45 &$-$21.09 &11.30 &0.032 & 35.51  \\
NGC 1572        &04 22 42.7 &$-$40 36 02 &$-$21.51 &11.16 &0.020 & 34.46  \\
NGC 1614        &04 33 59.8 &$-$08 34 44 &$-$21.19 &11.41 &0.016 & 33.96  \\
IRAS 05189-2524 &05 21 01.4 &$-$25 21 45 &$-$20.59 &11.89 &0.042 & 36.11  \\
ESO 255-IG007   &06 27 23.1 &$-$47 10 44 &$-$16.88 &11.67 &0.039 & 35.89  \\
NGC 2623        &08 38 24.0 &$+$25 45 17 &$-$21.20 &11.47 &0.018 & 34.37  \\
IRAS 08572+3915 &09 00 25.4 &$+$39 03 54 &$-$      &11.99 &0.058 & 36.85  \\
UGC 4881        &09 15 55.5 &$+$44 19 49 &$-$20.90 &11.57 &0.040 & 36.01  \\
UGC 5101        &09 35 51.4 &$+$61 21 11 &$-$20.78 &11.90 &0.039 & 36.00  \\
MCG +08-18-012  &09 36 30.7 &$+$48 28 10 &$-$20.16 &11.19 &0.026 & 35.07  \\
IC 563/IC 564   &09 46 20.3 &$+$03 02 43 &$-$21.58 &11.10 &0.020 & 34.50  \\
NGC 3110        &10 04 01.9 &$-$06 28 29 &$-$21.77 &11.10 &0.016 & 34.12  \\
IC 2545         &10 06 04.5 &$-$33 53 03 &$-$20.78 &11.57 &0.034 & 35.64  \\
IRAS 10173+0828 &10 20 00.0 &$+$08 13 35 &$-$19.11 &11.68 &0.048 & 36.42  \\
NGC 3256        &10 25 51.8 &$-$43 54 09 &$-$21.57 &11.44 &0.009 & 32.67  \\
IRAS 10565+2448 &10 59 18.1 &$+$24 32 34 &$-$21.76 &11.87 &0.042 & 36.20  \\
Arp 148         &11 03 53.9 &$+$40 51 00 &$-$      &11.50 &0.034 & 35.76  \\
MCG +00-29-023  &11 21 12.2 &$-$02 59 03 &$-$20.14 &11.36 &0.024 & 34.98  \\
IC 2810/UGC 6436&11 25 45.0 &$+$14 40 36 &$-$20.12 &11.50 &0.034 & 35.69  \\
NGC 3690        &11 28 31.9 &$+$58 33 45 &$-$22.04 &11.72 &0.011 & 33.38  \\
IRAS 12112+0305 &12 13 45.7 &$+$02 48 39 &$-$20.21 &12.19 &0.072 & 37.32  \\
%Mk 231         &12 56 14.2 &$+$56 52 28 &$-$22.08 &12.32 &0.042 & 36.15  \\
ESO 507-G070    &13 02 52.1 &$-$23 55 19 &$-$21.28 &11.31 &0.021 & 34.63  \\
UGC 8335        &13 15 29.2 &$+$62 07 11 &$-$20.22 &11.60 &0.031 & 35.51  \\
UGC 8387        &13 20 35.3 &$+$34 08 22 &$-$20.38 &11.52 &0.023 & 34.91  \\
NGC 5256        &13 38 17.7 &$+$48 16 34 &$-$21.80 &11.37 &0.027 & 35.31  \\
NGC 5257/5258   &13 39 53.1 &$+$00 50 22 &$-$21.83 &11.37 &0.022 & 34.80  \\
Mk 273          &13 44 42.1 &$+$55 53 13 &$-$20.22 &12.10 &0.038 & 35.92  \\
NGC 5331        &13 52 16.5 &$+$02 06 09 &$-$20.72 &11.43 &0.033 & 35.62  \\
Arp 302         &14 57 00.3 &$+$24 36 56 &$-$20.84 &11.59 &0.034 & 35.67  \\
Mk 848          &15 18 06.3 &$+$42 44 37 &$-$20.41 &11.72 &0.040 & 36.07  \\
IRAS 15250+3609 &15 26 59.4 &$+$35 58 38 &$-$21.21 &11.89 &0.053 & 36.74  \\
Arp 220         &15 34 57.3 &$+$23 30 12 &$-$20.78 &12.12 &0.018 & 34.37  \\
NGC 6090        &16 11 40.3 &$+$52 27 26 &$-$21.51 &11.35 &0.029 & 35.42  \\
IRAS 16164-0746 &16 19 11.8 &$-$07 54 03 &$-$20.51 &11.40 &0.027 & 34.93  \\
NGC 6240        &16 52 58.9 &$+$02 24 03 &$-$21.68 &11.85 &0.024 & 34.96  \\
IRAS 17208-0014 &17 23 21.9 &$-$00 17 00 &$-$20.13 &12.30 &0.043 & 36.18  \\
IC 4687/86      &18 13 39.6 &$-$57 43 31 &$-$20.60 &11.35 &0.017 & 34.12  \\
IRAS 18293-3413 &18 32 40.2 &$-$34 11 26 &$-$19.50 &11.63 &0.018 & 34.28  \\
NGC 6926        &20 33 06.2 &$-$02 01 40 &$-$22.15 &11.11 &0.020 & 34.52  \\
NGC 7130        &21 48 19.5 &$-$34 57 09 &$-$21.09 &11.21 &0.016 & 33.96  \\
\hline

\end{tabular}
\end{center}
\end{table*}

%---------------------------------------------------------------------------
\section{Observations}

The galaxies in Table~\ref{tab:gallist} were observed 
between October 1999 and October 2001 using
three different telescopes: the {\em Telescopio Nazionale Galileo} (TNG)
equipped with the NIR camera ARNICA (Lisi et al., 1993) until September 2001
and NICS (Baffa et al., 2001) afterward, 
the ESO NTT telescope with SOFI (Lidman et al., 2000)
and the Kuiper/Steward 61\arcsec\ infrared telescope (AZ61) on Mount Bigelow,
equipped with a near-IR NICMOS3 camera.

In all cases we used the local realization of the ``K short'' filter
(K$^\prime$),
similar to K but excluding the thermal part of the spectrum above 2.3$\mu$m.

The sampling time was chosen
in order to maximize the number of detections. At the limiting distance of 
our sample, 200 Mpc, an average SN remains above the detection limit for
about 40 days, increasing to 110 days at 100 Mpc. To be
conservative
and concentrate the observations around nights with a full moon,
we chose to observe the sample every 30 days. The final sampling is not very
uniform as the atmospheric conditions and the actual scheduling on the three
telescopes introduced some distortions on the original plan.
We collected 234 observations, with an average number of 5.1 observations per
galaxy (see Table~\ref{tab:ct}).

The typical on-source integration times were about 20-30 min for NTT and TNG
and 45-60 min for AZ61. Typical values of the seeing FWHM
were 1.2 arcsec at TNG, 1.0 arcsec at NTT and 1.5 arcsec at AZ61.
The photometric calibration was achieved by using the Hunt et al. (1998) 
and Persson et al. (1998) standard stars.

%---------------------------------------------------------------------------
\section{SN detection method, limiting magnitudes, and the SNe detections}
\label{sec:limits}

The comparison of the images of the same galaxies was performed with ISIS,
a tool developed by Alard \& Lupton (1998), refined by Alard (2000)
and available at http://www.iap.fr/users/alard/package.html. First,
all the images of a galaxy were aligned by using field stars or the 
galaxy nucleus. Second, the image with the best point-spread-function (PSF)
was identified and used as reference for the comparison with the 
other images.
In order to minimize the residuals without losing spatial resolution,
the PSF of the reference image was convolved with an appropriate kernel 
determined, for each image pair, by a linear fit.
ISIS was found to give better results than other similar tasks, 
such as the IRAF task ``immatch''.
Finally, the images were normalized to the same total flux and subtracted.

On average ISIS is able to subtract 97\% of the total galaxy flux, with a mode
of 98\%. The residuals were usually concentrated
at the position of the galactic nuclei,
where the emission has a strong radial gradient.
Non circular structures such as the 
diffraction spikes on the spider arms are not reproduced by the ISIS kernel.
The presence of these residuals makes the detection of nuclear 
SNe much more difficult, as discussed below.

\begin{table}
\caption[]{SN detection limits}
\label{tab:limits}
\begin{center}
\begin{tabular}{l c c}
\hline
Instrument & \multicolumn{2}{c}{5$\sigma$ limit K$^\prime$ mag} \\
           & off-nucleus  & on-nucleus \\
\hline
NTT/SOFI   & 19.3$\pm$0.4 & 16.9  \\
TNG/NICS   & 18.8$\pm$0.5 & 16.6  \\
TNG/ARNICA & 18.2$\pm$0.3 & 16.0  \\
AZ61       & 17.2$\pm$0.2 & 13.8  \\
\hline
\end{tabular}
\end{center}
\end{table}

Outside the nuclear regions, i.e., about 1 arcsec away from the galaxy
center, the SN detection limit was estimated through simulations
by adding point sources to the subtracted image
and trying to recover them. The detection was
performed by using SExtractor (Bertin \& Arnouts, 1996) with parameters set to
have secure detections of point sources (2.5$\sigma$ per pixel over an
area of 6 pixels, corresponding to about a 6$\sigma$ detection). 
The limiting magnitudes, defined at a completeness level of 50\%,
are listed in Table~\ref{tab:limits} for each telescope, together
with their spread because of variable seeing, transparency and integration
time. These values are in good agreement (within 0.1 mag) with the expectation
from the parameters of the detector and of the exposures.

\begin{figure}
	\centering
   	\includegraphics[width=9cm]{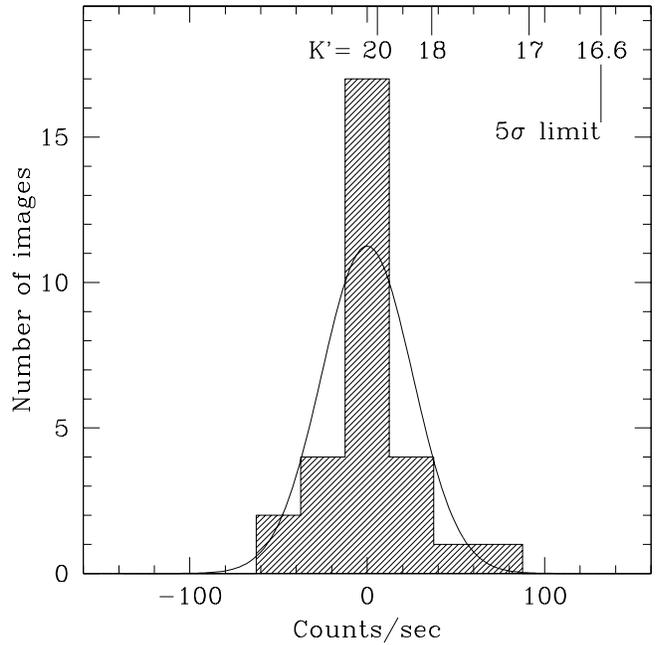}
   	\caption{Histogram of the nuclear residuals in the TNG/NICS
	images. The residuals are plotted in counts/sec and the 
	scale near the top shows the corresponding magnitudes. The gaussian fit
	and the 5$\sigma$ limit
	of K$^\prime$=16.6 are also shown.
	}
	\label{fig:maglim}
\end{figure}

In the central arcsec, the presence of the PSF subtraction
residuals makes it possible to
detect only brighter SNe. To estimate this limit we measured the
residuals inside a 1.5 arcsec aperture on each subtracted image, as
shown  in Fig~\ref{fig:maglim}. The magnitude corresponding to 
5$\sigma$ positive counts is taken as the detection limit. 
This is actually an upper limit to the real sensitivity as
in doing this we assume than the measured residuals are instrumental and
no SNe are actually present.

The analysis of the data revealed 4 SNe, listed in Table~\ref{tab:sne}.
Two objects, SN~1999gd and SN~2000bg, were independently
detected by us but they
had been already discovered by optical searches a few days before. 
Their K-band magnitudes
are listed in Table~\ref{tab:kmag}. The other two SNe, discovered in our
images, are presented in Maiolino et al., (2002). 
All the SNe were discovered out of the galactic nuclei;
the event closest to the nucleus of its host galaxy is
SN1999gw at 3.5\arcsec of distance.
As discussed in Maiolino et al. (2002), SN2001db is highly 
extincted ($A_V\sim5.6$), while the available data do not allow 
to constrain the extinction of the other SNe.

\begin{table}
\caption[]{Detected SNe}
\label{tab:sne}
\begin{center}
\begin{tabular}{c c c c c c}
\hline
SN     & Type & Galaxy   &\multicolumn{2}{c}{Discovery}& ref \\
       &      &          &  Date      &  Band        &   \\
\hline
1999gd & Ia   & NGC 2623 &1999 Nov 24 &   Opt.         & 1,2 \\
1999gw & II   & UGC 4881 &1999 Dec 16 &   K$^\prime$   & 3,4 \\
2000bg & IIn  & NGC 6240 &2000 Apr 01 &   Opt.         & 5,6 \\
2001db & II   & NGC 3256 &2001 Jan 09 &   K$^\prime$   & 3,7 \\
\hline
\end{tabular}
\end{center}
References. 
1: Li, 1999; 
2: Filippenko \& Garnavich, 1999;
4: Cresci et al., 2002.
3: Maiolino et al., 2002; 
5: Sato \& Li, 2000;
6: Jha \& Brown, 2000;
7: Maiolino et al., 2001;
\end{table}

\begin{table}
\caption[]{K$^\prime$ light curves of SN~1999gd and SN~2000bg}
\label{tab:kmag}
\begin{center}
\begin{tabular}{c c}
\hline
Date & K$^\prime$ \\
\hline
\multicolumn{2}{c}{SN 1999gd}\\
\hline
1999 Dec 15 & 16.0$\pm$0.2 \\
2000 Jan 25 & 17.4$\pm$0.2 \\
2000 Feb 11 & $>$18.0      \\
\hline
\multicolumn{2}{c}{SN 2000bg}\\
\hline
2000 Mar 14 & $>$ 17.0 \\
2000 Apr 10 & 16.0$\pm$0.3 \\
2000 May 12 & 16.0$\pm$0.3 \\
\hline
\end{tabular}
\end{center}
\end{table}

%---------------------------------------------------------------------------
\section{Computing the infrared SN rate}

This paper is devoted to computing the near-IR SN rate 
in starburst galaxies
by comparing the number of detections with the number of expected events.
To do this we need a few ingredients which will be the
subject of the following sections:
\begin{itemize}
\item the computation of the expected number of detections is based
on the estimate of the {\em control time}, i.e., the
amount of time than a SN in a given galaxy remains
above the detection limit (as defined by Zwicky, 1938). 
Therefore it is critical to have a good knowledge of the NIR light curve,
as discussed in section~\ref{sec:lc};
\item the expected number of exploding core-collapse SNe is also related to
the total SFR and therefore is expected to be proportional to the FIR
luminosity of the target galaxies. This proportionality
is discussed in section~\ref{sec:firsnr};
\item type Ia SNe are usually assumed to have old progenitors and therefore 
to be more related to the total mass in stars of the galaxies
than to the current or recent SFR. This might not be true for all types of
Ia SNe as discussed by several authors. In section~\ref{sec:Ia} we discuss 
the relation between type Ia SN rate and stellar mass along the Hubble
morphological types sequence. We will discuss the effect 
of including the type Ia SNe on the infrared SN rate;
\item very different detection limits apply to nuclei and to the rest of the
galaxies; as a consequence, the SN rates depend of the fraction of starburst
activity contained in the inner few arcsec of the galaxies. 
In appendix~\ref{sec:firdimen}
we discuss the existing FIR data, concluding that they do not really
constrain the size of the starbursts, even if in many cases
most of the star formation activity is concentrated in the
nuclear region;
\item section~\ref{sec:snr} contains the actual computation of the SN rate.
This is done considering three scenarios for the spatial dimension of the 
starbursts: we consider the two limiting cases that the fraction of starburst
in the central 2 arcsec is (a) negligible, and (b) dominant, and an
intermediate case in which 80\% of the starburst is contained in the
nucleus.
The rates are also computed both by 
considering the core-collapse SNe only 
and also by including the type Ia SNe.
\end{itemize}

%---------------------------------------------------------------------------
\begin{figure*}
	\centering
   	\includegraphics[width=\textwidth]{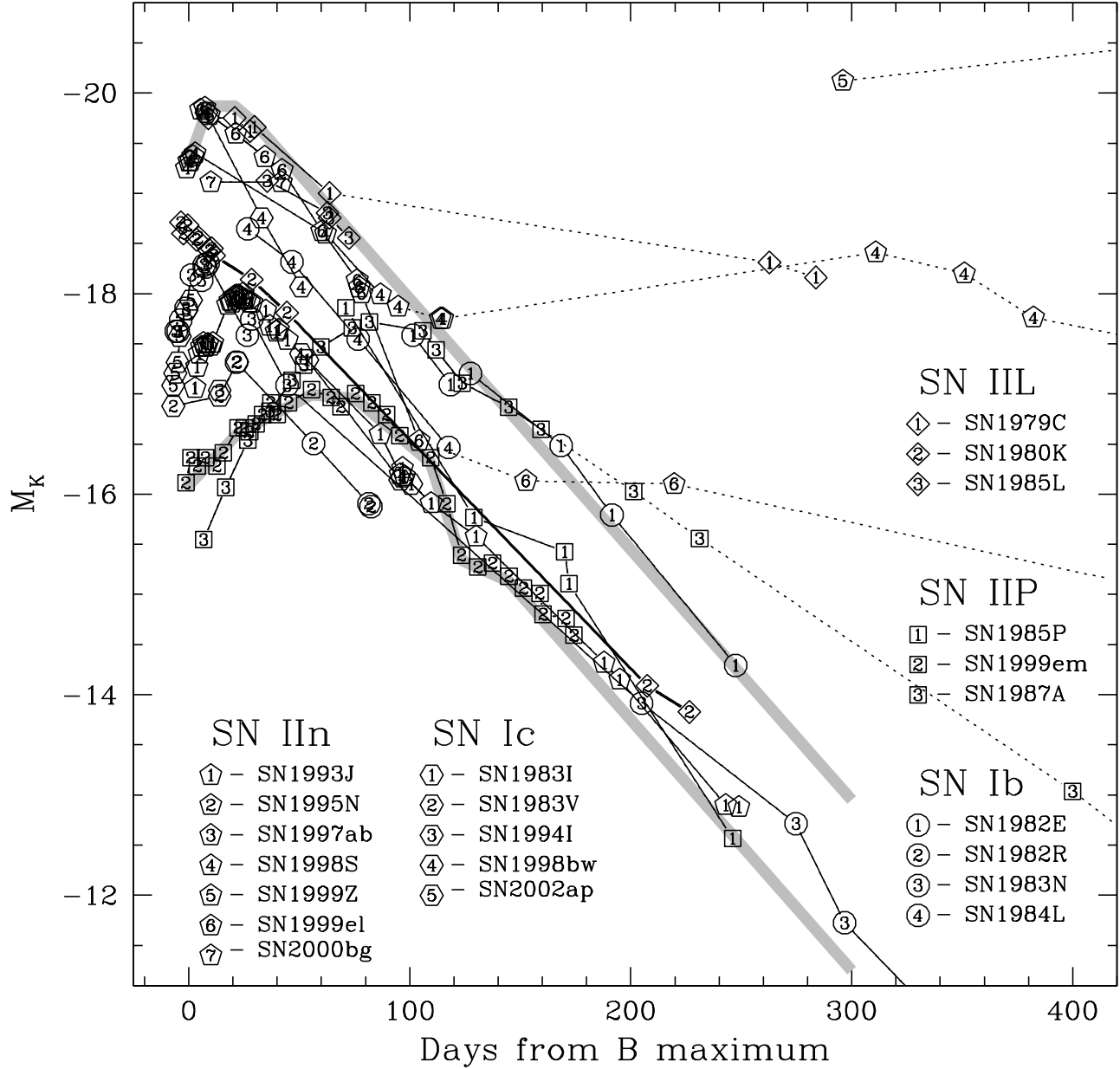}
   	\caption{K-band light-curves of all the core-collapse supernovae
	available in the literature. The upper- and lower-envelope
	used for the computation of the control time are shown
	by the thick grey lines. The thick solid line shows SN1980K.
	The part of the curves connected with dotted lines are varying too
	slowly to produce a detection and therefore were not considered
	to compute the envelopes. Some objects lie outside of the
	plot because only observations at epoch$>$400 days are available.}
	\label{fig:envelope}
\end{figure*}
 
\section{The K-band light curve of the core-collapse SN}
\label{sec:lc}

The evolution of the K band luminosity of the core-collapse SN
is currently not very well constrained. Mattila \& Meikle (2001) made 
a compilation of the 12 curves available at that time in the K band
to derive the average light curve. 
The small number of available measures is due to the fact that NIR 
observations are not
available for all the SNe and that only events with a well known epoch of
the optical maximum are considered.
They divided the
core-collapse supernovae in two classes, the ``ordinary events'' showing
decline rate similar to the optical band, and the ``slow declining events''
with NIR luminosity almost constant for hundreds of days after the optical
maximum. We have added 9 new light curves for a total number 
of more than 200 NIR observations (see Table~\ref{tab:lc} and references 
therein).
We have also included SN~1987A which, in the near-IR, has
a less peculiar behavior than in the optical.

The resulting sample is not complete, so it should not be used for statistics.
Nevertheless the fraction of the slow declining events appears to be not
negligible: 7 out of 22 SNe show a peculiar behavior at some point
of their evolution, and 5 of them are classified as type IIn SNe from the
optical spectra. The fraction is even higher, 7 out of 12,
if only the SN with observations after 4 months from the maximum
are considered. Such a high fraction is certainly due to an observational
bias,
as SNe with NIR excess are more often observed at such late times (see,
for example, Gerardy et al. 2002). SN~1987A shows a peak luminosity
similar to the other events and a decline rate intermediate between the
ordinary and the slow declining events.

This brightening or stabilization of the luminosity
at NIR wavelengths is usually attributed to thermal emission from dust 
forming in the ejecta or present in the pre-existing circumstellar medium 
(e.g., Gerardy et al. 2002), whereas the heating could be due to the 
interaction between the ejecta and the circumstellar medium (Fassia et al.
2000, 2001; Di Carlo et al. 2002).
For this reason the effect is expected to be more common in
the events occurring in the dense star forming regions
as the type IIn SNe (Schlegel, 1990).
These SNe become much brighter than the others, but
their constant luminosity makes their detection by the usual ``blinking''
technique almost
impossible after the maximum light.

The 15 ``ordinary events'' show  an
inhomogeneous behavior during the first 4 months.
Most of them reach a maximum 10-20 days after the optical peak, and 
decline linearly afterward. 
The NIR luminosity peak of core-collapse SNe is similar to that of the
type Ia, estimated to be M$_K=-$18.8 25 days after the optical maximum 
(Meikle, 2000), even if 
the core-collapse SNe are usually much fainter in the optical.
SN~1999em is the least luminous event, with K magnitude about 1
mag fainter at the optical maximum than any other measured SN,
and shows a very delayed peak, occurring at about 70 days after the optical
maximum. 
The luminosity spread among the SNe of each class
is comparable with the spread among the different classes, i.e., 
there is no evidence
of systemic differences for the core-collapse SNe of different sub-types.

After the first 120 days, ``ordinary'' SNe show similar 
decline rates of about 0.025 mag/day, 
much faster than the optical decline rate of about 0.010 mag/day at 
comparable stages corresponding to the 
$^{56}$Co decay 
(0.0098 mag/day, e.g., Cappellaro \& Turatto 2000).

The spread of the absolute magnitude near the K-band peak is about 3 magnitudes,
and about 1.5 magnitudes at later times. To account this large
spread we have defined the upper and lower envelopes of the distributions (see
Fig.~\ref{fig:envelope} and Table~\ref{tab:envel}) 
and will use these curves to compute the maximum and minimum SN rate
expected from the SFR. The upper envelope is defined by SNe 1979C and 1982R,
the lower envelope by SN~1999em. A few SNe, such as 1983I and 1993J,
 are fainter than
1999em at epochs between 50 and 120 days, but are brighter near the optical
maximum and therefore were not considered for the lower envelope.

For the average light-curve we used SN~1980K, an event showing a linear decline 
and a peak magnitude near the average value. We preferred to use an observed 
event instead of an average between the two envelopes since the latter is
strongly dependent on
peculiar sub- or super-luminous events and does not correspond to any
observed SN.

\begin{table}
\caption[]{Core-collapse SNe used to define the NIR light curve.
Name, type, epoch of optical maximum, distance modulus, number of K band
observations and references are reported.}
\label{tab:lc}
\begin{center}
\begin{tabular}{l l c c c l}
\hline
SN     &Type & Epoch   &  (m-M)& N$_{\rm obs}$ & ref \\
       &     & (JD)    &       &      &   \\
\hline
1979C  & IIL & 2443979 & 31.03 &   8  & 1     \\
1980K  & IIL & 2444543 & 28.78 &  13  & 2     \\
1982E  & Ib  & 2445057 & 31.39 &   7  & 3     \\
1982R  & Ib  & 2445248 & 31.22 &   5  & 3,4   \\
1983I  & Ic  & 2445443 & 31.30 &   6  & 5     \\
1983N  & Ib  & 2445533 & 28.49 &  17  & 1,5   \\
1983V  & Ic  & 2445672 & 31.30 &   4  & 6     \\
1984L  & Ib  & 2445943 & 31.52 &   6  & 5     \\
1985L  & IIL & 2446227 & 31.51 &   5  & 1     \\
1985P  & IIP & 2446349 & 30.47 &   6  & 1     \\
1987A  & IIP & 2446849 & 18.55 & many & 7    \\
1993J  & IIn & 2449095 & 27.80 &  25  & 8$-$14  \\
1994I  & Ic  & 2449451 & 29.60 &   2  & 15      \\
1995N  & IIn & 2449539 & 32.07 &  14  & 16      \\
1997ab & IIn & 2450186 & 33.49 &   3  & 16      \\
1998S  & IIn & 2450890 & 31.30 &  17  & 1,16,17 \\
1998bw & Ic  & 2450944 & 32.96 &   4  & 18      \\
1999Z  & IIn & 2451219 & 36.61 &   3  & 16      \\
1999el & IIn & 2451572 & 32.53 &  14  & 16,19   \\
2000bg & IIn & 2451636 & 35.11 &   2  & 20      \\ 
1999em & IIP & 2451483 & 29.38 &  35  & 21      \\
2002ap & Ic  & 2453314 & 29.52 &   6  & 22$-$25 \\
\hline
\end{tabular}
\end{center}
References. 
  1:  Mattila \& Meikle (2001); 
  2:  Dwek et al. (1983);
  3:  Graham (1985);
  4:  Muller (1982);
  5:  Elias et al. (1985);
  6:  Clocchiatti et al. (1997);
  7:  Bouchet et al. (1989);
  8:  Calamai et al. (1993);
  9:  Odewahn et al. (1993);
  10: Lawrence et al. (1993);
  11: Romanishin (1993);
  12: Smith (1993);
  13: Kidger et al. (1993);
  14: Matthews et al. (2002);
  15: Grossan et al. (1999);
  16: Gerardy et al. (2002);
  17: Fassia et al. (2000);
  18: Patat et al. (2001);
  19: Di Carlo et al. (2002);
  20: this paper;
  21: Hamuy et al. (2001);
  22: Motohara et al. (2002);
  23: Yoshii et al. (2002);
  24: Mattila et al. (2002)
  25: Mannucci et al., in prep..
\end{table}

\begin{table}
\caption[]{Reference K-band light curve and lower- and upper-envelopes 
for core-collapse SNe. Epoch refers to the optical maximum.}
\label{tab:envel}
\begin{center}
\begin{tabular}{r c c c}
\hline
Epoch  & Lower & SN~1980K & Upper   \\
\hline
 -1~~  & $-$16.07 & $-$18.66 & $-$19.25 \\
  9~~  & $-$16.30 & $-$18.44 & $-$19.87 \\
 22~~  & $-$16.60 & $-$18.23 & $-$19.85 \\
 30~~  & $-$16.70 & $-$18.11 & $-$19.70 \\
 55~~  & $-$16.98 & $-$17.57 & $-$19.08 \\
 75~~  & $-$16.95 & $-$17.11 & $-$18.58 \\
109~~  & $-$16.32 & $-$16.34 & $-$17.73 \\
123~~  & $-$15.37 & $-$16.02 & $-$17.38 \\
144~~  & $-$15.14 & $-$15.54 & $-$16.85 \\
300~~  & $-$11.25 & $-$11.64 & $-$12.95 \\
\hline
\end{tabular}
\end{center}
\end{table}

%---------------------------------------------------------------------------
\section{SN rate and FIR luminosity}
\label{sec:firsnr}

To compute the number of expected SNe in our sample, we need
to discuss the relation between FIR luminosity and SN rate 
and the role of type Ia SNe. These will be the subject
of the next two sections.\\

Several authors have computed the SN rates 
as a function of the host galaxy morphological type.
These estimates are based on optical observations
of local galaxies having dust contents and SFRs much lower than the objects in
our sample. 

Rates are generally normalized to the luminosity of the parent galaxy, 
either the B luminosity or \lfir. 
The former normalization is applied when using
the SN unit (SNu), i.e., the number of SNe per century per 
10$^{10}L_\odot$ of B luminosity. 
This luminosity has an ambiguous physical meaning: in starburst galaxies
it is roughly proportional to the SFR, in quiescent galaxies it is related to 
the total mass in stars, and in dusty starbursts it is very sensitive to the
dust content and distribution. Here we will use the rate of 0.93$\pm$0.28 SNu
by Cappellaro et al. (1999, Tab. 4) for galaxies later than Sb and 
including the rates for types II and Ib/c.

\begin{figure}
	\centering
   	\includegraphics[width=9cm]{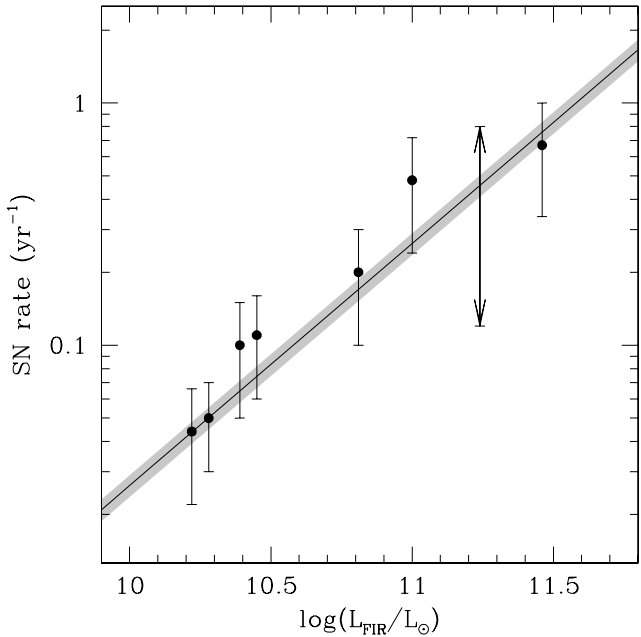}
   	\caption{SN rates vs. \lfir\ for a sample of galaxies
	with active star formation. From left to right:
	UGC~1347 (Hackenberg et al., 2000);
	NGC~253 (Mattila \& Meikle 2001 and references therein);
	NGC~7673 (Condon \& Yin, 1990)
	M~82 (Mattila \& Meikle, 2001 and references therein);
	NGC~4038 (Neff \& Ulvestad, 2000);
	NGC~1068 and NGC~7469 (Wilson et al., 1991).
	When not available, errors of 50\% of the measured values were assumed.
	The solid line shows a fit to these points
	passing through the origin, the shaded region the associated 
	1$\sigma$ error. The double arrow shows the upper- and lower limits for
	NGC~7771 (Smith et al., 1999) due to a poorly known correction for
	incompleteness.
	}
	\label{fig:snr}
\end{figure}

For galaxies with a significant SFR it is more useful, for many applications 
and especially when the core-collapse SNe are considered, to normalize 
the observed number of SNe to the
FIR luminosity, known to be proportional to the SFR. Several indirect
methods based on radio, FIR or mid-infrared observations have been
used to derive the SN rate in obscured starburst galaxies, as described in 
van Buren \& Greenhouse (1994) and
Mattila \& Meikle (2001).  These authors have derived the 
relation between the SN rate and \lfir\ using galaxies
of low FIR luminosity, and therefore a large extrapolation is needed
to apply these results to our sample. We have extended the observed range
toward high luminosities by using the relation between radio luminosity 
and SN rate in Condon \& Yin (1990) and data  
in Hackenberg et al. (2000)  and Wilson et al. (1991).
The SN rate appears to be proportional to \lfir\ with a small scatter
for luminosities between 1.6 and 30$\times10^{10}L_\odot$
as shown in Fig.~\ref{fig:snr},  without the detection of any dependence
on the luminosity of the galaxy or on the SFR.
Using the definition of \lfir\ in Eq.~(\ref{eq:fir}), we obtain the relation:

\begin{equation}
SNr = (2.4\pm0.1)\times(L_{FIR}/10^{10}L_\odot) \rm {~~SN/100yr}
\label{eq:snr}
\end{equation}

which, we stress, is based on indirect measures of the SN rate.
A similar value of the proportionality factor (2.7) was
also obtained by Mattila \& Meikle (2001) even though they use a slightly 
different definition of \lfir. In that case the data
range was below \lfir=10$^{10.8}L_\odot$, a factor of two below the faintest
galaxy in our sample. These authors discuss how this factor
is in good agreement with
the expectations from the models of galaxy stellar populations 
which predict the
FIR luminosity from the UV flux and the SN rate from the product of the
SFR times the integral of the Initial Mass Function (IMF)
over the appropriate mass range.
Finally, Cappellaro et al., (1999) obtained the ratio between
the rate of unobscured SNe
from optical monitoring and the FIR luminosity
of a large number of galaxies from the RC3 catalog.
They obtained a coefficient of 2.5,
in good agreement with our calibration. 

Smith et al. (1999) measured the SN rate 
in the central 10\arcsec\ of NGC~7771, a starburst galaxy
with \hbox{log(\lfir/$L_\odot$)=11.25.} This value cannot be used for the above relation
because it only referred to the central part of the galaxy.
To correct this SN rate to a value appropriate to the
total galaxy, we assume that the FIR luminosity is well traced by the 15$\mu$m 
emission (see, for example, Elbaz et al., 2002) and use the growth profile by
Dale et al. (2000). This is probably an overcorrection, as the size of
the region emitting 80\% of the 15$\mu$m luminosity is 54\arcsec\ while
Zink et al. (2000) obtain an upper limit to the 100$\mu$m emitting region 
of about 20\arcsec. 
In Fig.~\ref{fig:snr} we show the upper- and lower-limits to the
SN rate in NGC~7771: the value predicted by Eq.~(\ref{eq:snr}) is within
the permitted range.

%---------------------------------------------------------------------------
\section{The role of type Ia SNe}
\label{sec:Ia}

\begin{figure}
    \centering
    \includegraphics[width=9cm]{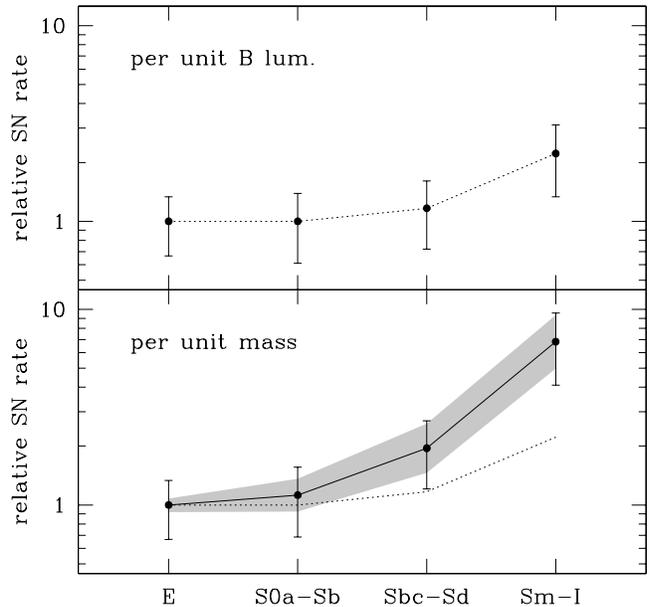}
    \caption{
    {\em Upper panel:} type Ia SN rate by Cappellaro et al. (1999)
    normalized to that in the elliptical galaxies.
    The rates were measured in SNu, i.e., were compared to the B 
    luminosity of the galaxies. The bars show the errors in the observed rates.
    {\em Lower panel:} the relative SN rate are normalized to the total 
    stellar mass of the galaxies as in Bell \& de Jong (2001).
    A factor of 7 of increase is observed from the ellipticals to the
    irregular galaxies. The grey region show the effect of the uncertainties 
    on M/L due to the color spread in each galaxy Hubble type. The dotted line
    is the same of the upper panel and is shown as reference.
    }
    \label{fig:snIa}
\end{figure}

The progenitors of the type Ia SNe are expected to be old systems
containing a white dwarf accreting mass from a companion star. 
If all Ia SNe are produced by the same old progenitors, 
their properties (as absolute magnitude at maximum, decline rate and expansion
velocity of the ejecta)
should not correlate with the current SFR of their host location.  
On the other hand several authors 
have discovered systematic differences
between type Ia SNe in galaxies of different Hubble types 
in terms of optical spectral properties
(Branch \& van den Berg, 1993, and references therein),
photometric differences (Della Valle \& Panagia, 1992),
photometric evolution (Hamuy et al., 2000)
and rates of occurrence (Della Valle \& Livio, 1994).
% We will discuss this last issue in greater detail
% in a forthcoming paper, where we show that type Ia SN rate
% exhibits a sharp increase from early- to late-type galaxies.

If the progenitors are old, the number of type Ia SNe
should also roughly correlate with the total stellar mass.
This fact is barely detectable from the SN rate expressed in SNu as
the B luminosity is not a good tracer of stellar mass.
On the contrary, the luminosity in the K band is a much better tracer,
as discussed for example by Brinchmann \& Ellis (2000).
The dominating uncertainties are due to the age of the stellar population,
and Bell \& de Jong (2001) propose a method to correct for this effect 
from the observed colors. We have therefore corrected the rates in 
Cappellaro et al. (1999)
by using the average B-K colors for galaxies of the various Hubble type 
by Fioc \& Rocca-Volmerange (1999) and the corrections to the M/L ratio
by Bell \& de Jong (2001). 
show an increase of about a factor of 2 of the SN rate in the
Sbc-Sd galaxies over that in the ellipticals, 
rising to about a factor of 7 for Sm-I galaxies.
These values could be affected by systematic errors, but the results are 
consistent with the hypothesis of a substantial enhancement of the type Ia SN
rate due to young stars.
The details of this computation and the the relation between the 
stellar mass and the type Ia rate will be subject
of a forthcoming paper.

Based on this kind of evidences, Della Valle and Livio (1994) 
found that late type spirals are more prolific producers of type Ia SNe
and proposed the association of part of the type Ia
SNe with younger stellar populations.
% It should be noted that an increase of type Ia SN is expected in 
% starburst galaxies even in the standard white dwarf scenario,
% as binary systems of two massive stars are expected to form, evolve
% rapidly and give rise to a type Ia SN.

For these reasons, we will make the calculations under two opposite limiting
hypothesis that (1) none, and (2) all the type Ia SNe derive from young 
progenitors. In the second hypothesis the expected SN rate is raised
by about 30\% as this is the fraction of type Ia SNe vs. the core-collapse SN
measured by Cappellaro et al. (1999) for galaxies later than Sbc.
In this case the number of detected SN is also increased, as the
type Ia SN~1999gd must be included in the sample, and the two effects
(larger expected rates and more observed events) tend to
cancel each other.

%---------------------------------------------------------------------------
\section{The infrared SN rate}
\label{sec:snr}

\begin{table}
\caption[]{Control time (days) for each galaxy for 
core-collapse (cc) and type Ia SNe using the off-nuclear limiting magnitude}
\label{tab:ct}

\begin{center}
\begin{tabular}{lccc}
\hline
Galaxy & N obs. & CT(cc) & CT(Ia) \\
       &        & (days) & (days) \\
\hline 
NGC~34         &    6  &  457.1 &   332.3  \\
NGC~232        &    3  &  399   &   265.5  \\
MCG+12-02-001  &    4  &  325.5 &   260.1  \\
IC~1623        &    5  &  416.6 &   287.2  \\
UGC~2369       &    5  &  338.1 &   244.4  \\
IRAS0335+15    &    8  &  324.7 &   211.4  \\
MCG-03-12-00   &   4   &  300.6 &   199.2  \\
NGC~1572       &    5  &  467.2 &   343.8  \\
NGC~1614       &    13 &  746.4 &   583.9  \\
IRAS0518-25    &    4  &  265.4 &   184.2  \\
ESO255-IG007   &    6  &  346.8 &   258.4  \\
NGC~2623       &    12 &  575.0 &   459.8  \\
IRAS0857+39    &    5  &   43.9 &   27.8   \\
UGC~4881       &    7  &  181.2 &   121.6  \\
UGC~5101       &   7   &  181.1 &   122.3  \\
MCG+08-18-012  &    2  &   77.1 &   61     \\
IC~563/IC564   &   3   &  101.1 &   70.1   \\
NGC~3110       &   6   &  405.9 &   305.2  \\
IRAS1017+08    &    5  &  184.8 &   145.9  \\
NGC~3256       &    5  &  582.4 &   458.4  \\
IRAS1056+24    &    8  &  161.0 &   112.6  \\
Arp~148        &   4   &  181.0 &   94.8   \\
MCG+00-29-023  &    6  &  325.4 &   248    \\
IC~2810/UGC6436&   4   &  188.7 &   125.6  \\
NGC~3690       &    9  &  555.1 &   442.7  \\
IRAS1211+03    &    5  &  102.6 &   83.4   \\
ESO507-G070    &    2  &  208.5 &   143    \\
UGC~8335       &   5   &  297.1 &   194.5  \\
UGC~8387       &    4  &  221.3 &   149.5  \\
NGC~5256       &    2  &   62.3 &   53.6   \\
NGC~5257       &    6  &  360.0 &   273.1  \\
Mk~273         &   4   &  131.3 &   77.1   \\
NGC~5331       &    4  &  242.8 &   149.3  \\
Arp~302        &    5  &  215.4 &   115.8  \\
Mk~848         &    3  &  116.4 &   70     \\
IRAS1525+36    &    2  &    0.0 &   0      \\
Arp~220 	   &    12 &  591.9 &   478.4  \\
NGC~6090       &    4  &  212.1 &   164.7  \\
IRAS16164-0746 &   2   &  195.2 &   115.3  \\
NGC~6240       &    10 &  459.3 &   380.9  \\
IRAS17208-0014 &   5   &  125.6 &    75.8  \\
IC~4687/86     &    3  &  512.1 &   350.4  \\
IRAS1829-34    &    3  &  503.8 &   325.8  \\
NGC~6926	   &    4  &  202.6 &   143.2  \\
NGC~7130       &    3  &  436.2 &   313.2  \\
\hline
\end{tabular}
\end{center}
\end{table}

The rates we can finally compute 
depend on the control time and therefore on the limiting magnitude.
In the central 1 or 2 arcsecs of the galaxy nucleus
the capability of detecting SNe is dimmed because of the presence of 
residuals after 
the subtraction, as discussed in Sect.~\ref{sec:limits}. 
Although the area affected by the residuals is a small fraction of the total
galaxy, this part could contain a significant fraction of
events because it could host a compact nuclear starburst region at the 
bottom of the potential well. 
In appendix~\ref{sec:firdimen} we discuss the information on the
size of the starburst region coming from the mid-IR and far-IR data, 
and we conclude that
many galaxies might be dominated by a compact starburst but that
the spatial resolution of the data is too poor to obtain definitive conclusions.

As a consequence we compute the rates in the
two limiting cases that the FIR flux in the nuclear region
is (a) negligible (labeled as {\em off-nuclear} case) and (b) dominant
({\em nuclear}). 
Given that in many cases most of
the starburst activity seem to originate  in the nuclear part of the galaxy,
we have also considered a third intermediate case ({\em 0.2off+0.8nuc})
in which 80\% of the starburst is contained in nucleus and the remaining 20\%
is distributed in the rest of the galaxy.
Such a scenario is conservative but compatible 
with all the observations mentioned in appendix~\ref{sec:firdimen}. 

The rates are computed both for core-collapse SNe only 
and also by including the type Ia events. The results are summarized in
Table~\ref{tab:rates}.

In each case, we have computed the control time
(i.e., the amount of time a SN remains visible in our images)
for each galaxy in Table~\ref{tab:gallist} by using the 
average type Ia light curve and, for the core-collapse events,
the light curve of SN~1980K and the upper- and lower envelopes in
Fig.~\ref{fig:envelope}. From the control time we computed
the expected number of detected SNe using the galaxy's B and FIR luminosities
and using the appropriate limiting magnitude
(see discussion in Sect.~\ref{sec:limits}).
The control times computed in this way 
corresponding to the off-nuclear limits
are listed in Table~\ref{tab:ct}.
Table~\ref{tab:rates} gives the expected number of
detections for the whole sample 
and the corresponding number of detections.

The uncertainties on the number of expected events
are due to (1)
the spread of the SN K-band luminosity 
(see Fig.~\ref{fig:envelope}), 
(2) the uncertainties in the ratio between B luminosity
and SN rate, and (3)
the spread of the limiting magnitude in our images. 
The first effect is the dominant one:
for example, when using the average light curve (SN~1980K) and
the off-nuclear limits, 0.70 core-collapse SNe
are expected from the B luminosity, while 0.90 and 0.46 events
are expected when using the light curves corresponding, respectively,
to the upper and lower envelope of Figure~\ref{fig:envelope}.

From Table~\ref{tab:rates} it is apparent that in all cases
the number of SNe predicted from 
the B luminosity, between 0.31 and 0.84,
is much lower than our 4 detections.
When the SN rate is estimated from the
FIR luminosity we obtain drastically different results: 
the expected number of SNe is higher by a factor of 3$-$10
than the SNe actually detected
(see next section).  \\

The number of the detected SNe can also be used to
derive a {\em near-IR SN rate} (\snrnir),
i.e., based on near-IR observations. 
We consider only the case
with 80\% of FIR luminosity coming from the nucleus.
The rates were computed by dividing the total number of SNe 
by the sum of the luminosity 
of the galaxies times the control time. 
By using both the B luminosity and \lfir\ we obtain, respectively:
\begin{equation}
SN^{NIR}_r = 7.6 ~~\pm~3.8~\pm~2.8~~~SNu~~(B)
\end{equation}
\begin{equation}
SN^{NIR}_r = 0.53~\pm~0.27~\pm~0.21~~~SNuIR
\label{eq:snr}
\end{equation}
The two errors terms in each equation are due to
the statistics of the number of detections 
and the uncertainties in the computation of the
control time, respectively. 
It should be noted that the first one (statistics) is the dominant term
due to the small number of detections, 
and that the second one (control time) is not a standard 
``1$\sigma$'' 68\% value but rather a total 100\% uncertainty.

\begin{table}
\caption{Number of observed SNe compared with the number of expected events
as derived from the B and FIR luminosities.
cc indicates core-collapse SNe only, cc+Ia includes also type Ia SNe.}

\label{tab:rates}
\begin{center}
\begin{tabular}{lr|cc}
\hline
              &                  &     cc        & cc+Ia \\
\hline
Detected SNe  &                  &      3        &   4          \\
\hline
                                 &{\em off-nuclear} &0.70$\pm$0.22 &0.84$\pm$0.23 \\
\makebox[2.2cm][l]{Expec. from $L_B$}&{\em  nuclear}&0.25$\pm$0.22 &0.31$\pm$0.23\\
                                 &{\em 0.2off+0.8nuc}&0.34$\pm$0.18 &0.49$\pm$0.18\\
\hline
                                 &{\em off-nuclear } & 35$\pm$16    & 42$\pm$17   \\
\makebox[2.2cm][l]{Expec. from \lfir}&{\em  nuclear} & 10$\pm$8     & 11$\pm$8    \\
                                 &{\em 0.2off+0.8nuc}& 15$\pm$7     & 18$\pm$7    \\
\hline
\end{tabular}
\end{center}
\end{table}

%---------------------------------------------------------------------------
\section {Comparison between expected and detected SN rates}
\label{sec:stat}

The first conclusion is that NIR searches for SN in starburst galaxies
are, as expected, more efficient than similar searches at optical wavelengths
since at least some of the SNe occurs in dusty environments.
Two SNe were in fact only detected in the NIR, and
Maiolino et al. (2002) demonstrated that at least SN~2001db was too absorbed
($A_V\sim5.6$ mag)
to be detected during routine optical SN surveys even at its maximum. 
This is also the case for the
type Ia SN~2002cv discovered in the NIR by Di Paola et al. (2002) 
and for which Meikle and Mattila (2002)
derived an extinction in excess of $A_V$=6 mag.\\

The second conclusion is that we are detecting more SNe than expected from the
B luminosity, both considering or neglecting the contribution from type Ia
SNe.  The hypothesis that the detected number of core-collapse SNe (3) 
is still compatible with the expectations (0.34) can be
rejected at more than the 99.5\% confidence level. 
A very similar result is obtained if we include the type Ia SNe 
(0.49 predicted vs. 4 observed).

This high value for the observed  SN rate 
reflects the higher extinction affecting
the B light (which is the normalizing factor of SNu)
of the galaxies in our sample,
and the higher SN detection efficiency of the NIR
observations with respect to the optical. 

This result is not only due to obscuration, as 2 of the 4 detected SNe
were also observed in the optical. For this reason the low success rate of 
previous NIR surveys for SNe is probably due to modest star forming activity
of the monitored galaxy sample: 
for this work we used galaxies with higher rates of star formation
which translates into a higher expected SN rate.\\

The third conclusion is that the majority of SNe expected from the FIR
luminosity are
still missing, i.e., we have detected only about 10$-$30\% of the expected SNe.
This small number of detections can be explained in several ways:
\begin{enumerate}

\item Most of the SNe are so embedded into
the dust that their luminosity is vastly reduced even at near-IR wavelengths.
The average extinction $A_V$ needed to reduce the expected 
number of 15 SNe to the observed number of 4 is $A_V=31$ mag,
or $A_V>26$ (corresponding to 8 SNe) with a formal confidence level of 90\%.
These values for the extinction are
similar to those obtained by Genzel et al. (1998) from mid-infrared
spectroscopy of several luminous galaxies, and by Mattila \& Meikle (2001) 
for M82 by assuming a gas-to-dust ratio similar to the Milky Way. In this
case, a NIR search for SN can detect only a fraction of the total SN rate, and
a radio survey should also be planned.

\item If 100\% of the FIR flux comes from the central arcsec, 
only 11 events are expected because of the residuals present in the nuclear
region even after applying a PSF matching algorithm. When considering also
the uncertainties in the predictions, dominated by the dispersion in 
luminosity of the SNe, the number of expected SNe is compatible
with the observed one without the need of a large additional
extinction. The presence of a large number of SNe in the central arcsec
of the starburst galaxies can be tested by near-IR monitoring from space:
NICMOS on HST can provide images with higher resolution ($\simeq0.2$\arcsec)
and the more stable PSF to considerably reduce the central residuals
and reveal the nuclear SNe. Early results on such a project based on 4
galaxies will be presented in a forthcoming paper (Cresci et al., 2003, in
preparation).

\item Another possibility is the presence of AGNs 
dominating the FIR flux of most of the galaxies:
in this case the FIR flux would not be related to the SN rate. 
There is evidence that the AGN does not
dominate the energetics of the galaxies with \lfir$<2\times10^{12}L_\odot$
(e.g., Sanders \& Mirabel, 1996). Nevertheless 
it is sometimes difficult to exclude the presence of an
AGN completely enshrouded by dust and therefore 
elusive in the optical/IR or even in the X-rays
(Marconi et al., 2000; Maiolino et al., 1998). 

\item Eq.~(\ref{eq:snr}) may overestimate the SN rate for
a given FIR luminosity. 
Many of the points in Fig.~\ref{fig:snr} and in particular that for the most
luminous galaxy NGC7496 are, directly or indirectly, derived from radio
observations and make use of the empirical relations by Condon \& Yin (1990).
In turn, this is based on radio observations of the Milky Way and on the 
estimate of its SN rate (Tammann, 1982). If the SN radio luminosity has a 
strong dependence on environment or if the estimates of the Galactic SN rate
are subject to strong incompleteness, the coefficient in Eq.~(\ref{eq:snr})
could be inadequate.
We believe that this is not a likely possibility as
there is a robust concordance among several different methods and
this relation has been stable through the years 
(see, for example, Rieke et al., 1980).

\item Part of the difference could also be due to a different IMF
in the starburst regions with respect to more quiescent galaxies.
Such an effect, studied in M82 by Rieke et al. (1993), could change
the relation between luminosity and SN rate. If
present, it is not likely to be a dominant effect 
as both the FIR luminosity and the
core-collapse SNe derive from massive stars.

\item An environment dependence could also be present in the SN luminosity: as
an example, SN in dense environments are expected to show a flatter light curve
because of the presence of a light echo (e.g., Roscherr \& Schaefer, 2000).
This has some consequences on the expected number of SNe:
events showing the ``ordinary'' decline rate can be detected even if
observed only after the maximum light, while the detection of
``slow'' events need the comparison of images taken
before and after the explosion.
In the former case the total useful time contains also one control time 
before the first image, in the latter case only the time between the images
should be considered, and the control time is reduced by about 25\%.
Therefore this effect could be present but is not strong enough to explain the
reduced number of detections.

\end{enumerate}

%-----------------------------------------------------------------------------
\section{Summary and conclusions}

We have monitored 46 infrared luminous galaxies at near-infrared wavelengths
to look for obscured SNe, and detected 4 SNe, 3 core-collapse and 1 type Ia
events. 
The analysis of this data set has allowed us to derive for the first
time the infrared SN rate in starburst galaxies.

To apply the control-time technique,
we have collected more than 200 observations of 21 core-collapse SNe
to derive their average, maximum and minimum K-band luminosities.
We have also 
extended the relation by Mattila \& Meikle (2001) 
between SN rate and FIR luminosity toward objects with higher SFR
in order to cover the luminosity range of the galaxies in our sample.
This relation remains linear, i.e., the SN rate is proportional to the
FIR luminosity, pointing toward a substantial 
uniformity in this property between quiescent and active galaxies.

We have computed the rate of type Ia SN as a function of the stellar mass
of the galaxy,
finding a clear evidence of its increase toward galaxies with higher
activity of star formation. This is an other indication that 
a fraction of type Ia SNe is probably be associated with relatively
young stellar populations.

We estimated the expected SN rate from the optical (B) and
FIR luminosities of the sample of galaxies and compared those predictions 
to the number of detections. 
The SN rate predicted from the B-band luminosity is several times 
below the observed rate. This is likely to be
explained by a number of reasons, the most important of which is
the effect of dust obscuration on the optical luminosity 
of the galaxies. On the contrary, a factor of 3$-$10 more SNe than observed
are predicted by the FIR luminosity. 
Among various possibilities, this can be explained by the presence of
dust extinction of at least A$_V=30$. This result prompts to the revision of
the local SN rate obtained in the past, as optical monitoring campaigns
have probably missed a significant fraction of SNe.

%-----------------------------------------------------------------------------
\begin{acknowledgements}
We are grateful to S. Mattila and P. Meikle for useful discussions 
and to Marcia and 
George Rieke for the use of the AZ61 near-infrared camera.
\end{acknowledgements}

%-----------------------------------------------------------------------------
\appendix

\section{The spatial dimension of the starburst activity}
\label{sec:firdimen}

The spatial resolution at 100$\mu$m is too low to image the 
spatial distribution of the starburst activity: the highest resolution achieved
of this type of galaxies is about 23\arcsec\ by Zink et al. (2000), much larger
that the scale of about 2\arcsec\ of importance here. 
Even at this coarse resolution, these authors were able to clearly
resolve 6 of their 22 galaxies, and obtain ``some evidence for extension''
in 7 more objects: as a result about two third of this incomplete sample
have 100$\mu$m dimension above 20\arcsec.

\begin{figure}
   \centering
   \includegraphics[width=9cm]{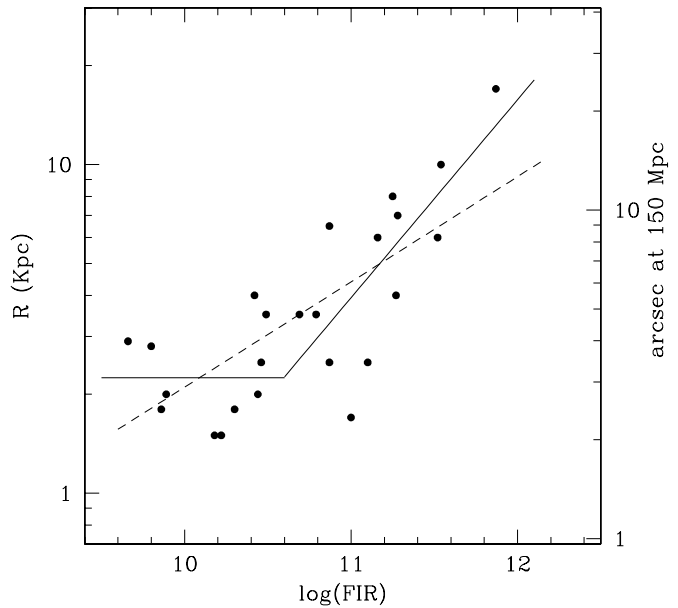}
   \caption{Size of the region emitting 80\% of the luminosity at
   15$\mu$m vs. the FIR luminosity. The data points are the galaxies present
   in both the Dale et al. (2000) and Soifer et al. (1987) catalogs.
   The dashed line shows the linear fit to all the points,
   while the solid lines are the linear fit for \lfir$>4\times10^{10}L_\odot$
   and the average below this limit. 
   }
   \label{fig:mirsize}
\end{figure}

Indirect evidence, but on a smaller scale, 
can be found by considering the images at mid-infrared wavelengths.
The tight correlation existing between the emission at 100$\mu$m and at 
15$\mu$m for galaxies over 4 orders of magnitudes of luminosity
(Elbaz et al., 2002)
implies that the mid-infrared should trace the 100$\mu$m emission
(unless an AGN is present).
High resolution mid-infrared images of luminous infrared galaxies
were obtained by Dale et al. (2000) using ISO.
Most of the galaxies show emission coming from several kpc of size,
corresponding to several (2-20) arcsec in our galaxies.
A possible correlation is also seen between \lfir\ and the size of the
region emitting 80\% of the 15$\mu$m emission (see Fig.~\ref{fig:mirsize}) 
for the galaxies with \lfir$>10^{10}L_\odot$. According to this relation, no 
starburst region in the galaxies in our sample should have dimensions
below 4\arcsec.

Bushouse et al. (1998) studied the relative distributions of several
indicators of star formation, such as H$\alpha$, mid-infrared, far-infrared
and radio continuum, and found a good correlation 
between these indicators in the central regions of a
few nearby galaxies. At 10$\mu$m, the emitting
source is usually not resolved at their resolution of about 5\arcsec,
indicating that most of the starburst is concentrated in the nucleus. 
The upper limits to the size of the far-infrared region are generally
compatible with those from the mid-infrared. In one
case, NGC~660 (log(\lfir/$L_\odot$)=10.38), the infrared emission 
is resolved both in the mid- and in the far-infrared: 
at 10$\mu$m Bushhouse et al (1998) found a size of about 320 pc,
while Smith \& Harvey (1996) obtained about 2 kpc at 100$\mu$m.  
At least in this
case the far-infrared emission seems more extended than at shorter
wavelengths.

Soifer et al. (2001) imaged a small number of high-luminosity galaxies
with Keck at 12.5$\mu$m with a field-of-view
of 17\arcsec$\times$17\arcsec. They generally find that most of the
emission comes from the central 2$-$3\arcsec, much more compact regions than
found in the sample by Dale et al. (2000). 
This result is somewhat expected because of the 
small field-of-view, but suggests once more that the nuclear star formation
contributes most of the mid- and far-IR.

In conclusions, current data don't put tight constraints of the dimension of
the starburst activity and therefore it is not possible to estimate the amount
of FIR flux originating the in the central 2 arcsec. 
For this reason we have computed the SN rate in the two limiting cases that
none and all the FIR flux comes from the nuclear region.
Given that most galaxies
seem to be dominated by nuclear starburst we have also computed the SN rate by
assuming the 80\% of the FIR flux originates from the nucleus and 20\%
from the rest of the galaxy.

\end{document}